# Critical Problems of Energy Frontier Muon Colliders: Optics, Magnets and Radiation

**Yu.I. Alexahin, E. Barzi, E. Gianfelice-Wendt, V. Kapin, V.V. Kashikhin,
N.V. Mokhov, I. Novitski, V. Shiltsev, S. Striganov, I. Tropin, A.V. Zlobin**

*Fermi National Accelerator Laboratory,
Batavia, IL 60510, USA*

## Executive summary

This White Paper brings together in a coherent form the results of the authors' previous studies on a Muon Collider (MC) and presents a design concept of the 6 TeV MC optics, the superconducting (SC) magnets, and a preliminary analysis of the protection system to substantially reduce radiation loads on the magnets as well as particle backgrounds in the collider detector. The SC magnets and detector protection considerations impose strict limitations on the lattice choice, hence the design of the collider optics, magnets and Machine Detector Interface (MDI) are closely intertwined. As a first approximation we use the Interaction Region (IR) design with $\beta^*=3$ mm, whereas for the arcs we rescale the arc cell design of the 3 TeV collider.

Magnets based on traditional cos-theta coil geometry $Nb_3Sn$ superconductor were used to provide realistic field maps for the analysis and optimization of the arc lattice and IR design, as well as for studies of beam dynamics and magnet protection against radiation. An important issue to address is the stress management in the coil to avoid substantial degradation or even damage of the brittle SC coils. Stress management concepts for shell-type coils are being experimentally studied for high-field accelerator magnets based on LTS ($Nb_3Sn$) and HTS (Bi2212 and ReBCO) cables.

In the assumed IR designs, the dipoles close to the Interaction Point (IP) and tungsten masks in each interconnect region (needed to protect magnets) help reducing background particle fluxes in the detector by a substantial factor. The tungsten nozzles in the 6 to 600 cm region from the IP, assisted by the detector solenoid field, trap most of the decay electrons created close to the IP as well as most of the incoherent $e+e-$ pairs generated in the IP. With sophisticated tungsten, iron, concrete and borated polyethylene shielding in the MDI region, the total reduction of background loads by more than three orders of magnitude can be achieved.

We believe that all the most critical concerns of the MC optics design, SC magnets, and radiation protection have not only been conceptually resolved but also addressed in sufficient detail for a 3-6 TeV Muon Collider, which can be considered either for the International Muon Collider Collaboration or as a Fermilab site filler.

Contents

**Introduction**

    A high-energy high-luminosity Muon Collider (MC) is considered as a most promising discovery machine [1]. Since the muon energy is not divided between the constituents like in the case of protons, the energy reach of a MC is an order of magnitude higher than that of a *pp* collider with nominally the same collision energy [2], [3]. We discuss the critical problems of an energy frontier Muon Collider with center of mass (c.o.m.) energy of 6 TeV based on previous design studies of lattice, magnets and their protection from radiation, detector backgrounds and radiation issues for muon colliders with c.o.m. energy from 125 GeV to 4 TeV [4]-[12]. The energy reach of this machine noticeably exceeds the $\sqrt{s} = 27$ TeV goal of the LHC energy upgrade making the 6 TeV MC a competitive option to explore this energy scale. There is also a good understanding of the challenges that such a machine may present [7]-[9] as well as a preliminary design of the Interaction Region (IR) for $\sqrt{s} = 6$ TeV MC [5], [10].

    Problems of muon generation, cooling and acceleration are not discussed in this paper. The muon beam parameters used in this paper are established by the U.S. Muon Accelerator Program (MAP) [13], [14]. The following topics will be addressed:
- MC lattice design
- High-field /high-gradient superconducting (SC) magnets

- Radiation protection and Machine-Detector Interface (MDI)

This White Paper summarizes in a coherent form the results of our previous studies on Muon Colliders and presents design concepts of the 6 TeV MC optics and SC magnets, and a preliminary analysis of the protection system to reduce radiation loads on the superconducting coils of the MC magnets as well as particle backgrounds in the collider detector. It will also refine further the problem of neutrino fluxes around MC and methods of its mitigation. The realistically achievable MC energy with the proposed concept and the required R&D effort will be presented and discussed.

1. **High Energy MC Lattices**

These are requirements that are either specific or more challenging in the case of a muon collider:
- Low beta-function at IP: $\beta^*$ values of a few millimeters are considered for muon colliders in the c.o.m. energy range 3-6 TeV.
- Small circumference $C$ to increase the number of turns (and therefore interactions) that the muons make during their lifetime.
- High number of muons per bunch: $N\mu \sim 2\times10^{12}$ and higher is envisaged.
- Protection of magnets from heat deposition and detectors from backgrounds produced by secondary particles.

Although short muon lifetime, ~2000 turns, creates a number of problems, such short time is not enough for high-order resonances to manifest themselves. Other sources of diffusion such as Intrabeam Scattering (IBS) or residual gas are also too weak to produce a halo. Thus if the muon beams are pre-collimated e.g. at $3\sigma$ before injection into the collider ring their distribution is likely to stay confined. This would relax the requirements on the dynamic aperture and on the efficiency of the halo removal from the ring.

In order not to lose much in luminosity due to the hourglass effect the bunch length should be small enough: $\sigma_z \leq \beta^*$. With longitudinal emittance expected from the final cooling channel this will render the momentum spread $\sigma_z/p \sim 0.001$ which is an order of magnitude larger than in hadron colliders such as the Tevatron or the LHC. Therefore, a high energy MC must have a large momentum acceptance and - to obtain small $\sigma_z$ with a reasonable RF voltage - low momentum compaction factor $|a_c| \sim 10^{-5}$.

Also, for beam energies above 2 TeV there is a peculiar requirement of absence of straight sections longer than ~0.5 m in order not to create "hot spots" of neutrino fluxes around the ring. As a consequence, quadrupoles must have a dipole component to spread background particles and direction vectors of the neutrino from muon decays. This creates difficulties with $\beta^*$-tuning sections which must allow for $\beta^*$ variation in a wide range without breaking the dispersion closure.

**1.1 SR lattice and IR design for 3 TeV MC**

The parameters of a Muon Collider with a c.o.m. energy of 3 TeV and an average luminosity of $4 \cdot 10^{34}$ cm$^{-2}$s$^{-1}$ are summarized in Table 1. The arc cell lattice and variations of horizontal and vertical beta-functions and dispersion are shown in Fig. 1 (top). The lattice consists of strong bending dipoles and combined-function quadrupoles with large dipole fields. Horizontal ($x$) and vertical ($y$) beam size variations ($4\sigma$) in the arc magnets are shown in Fig. 1 (bottom). The aperture of the magnets is determined by the following criterion $D_{x,y}=8\sigma_{max}$.

Table 1: MC Storage Ring Parameters.

| Parameter | Unit | Value |
| --- | --- | --- |
| Beam energy | TeV | 1.5 |
| Circumference | km | 4.5 |
| Momentum acceptance | % | ±0.5 |
| Transverse emittance, $\varepsilon_N$ | π·mm·mrad | 25 |
| Number of IPs | | 2 |
| $\beta^*$ | cm | 0.5 |

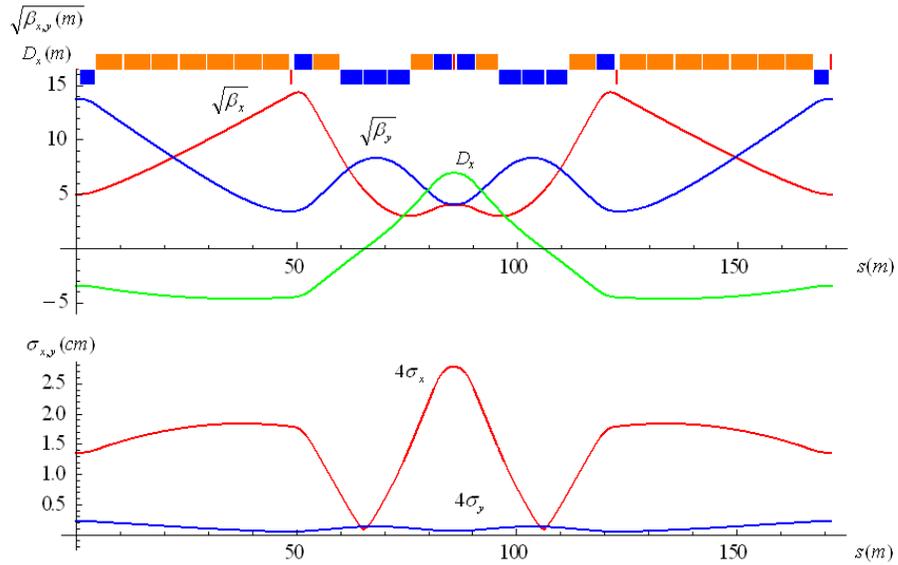

Figure 1: MC arc cell concept and beam size in magnets.

The target values of dipole field $B$, field gradient $G$ and beam aperture $D_{x,y}$ for the combined - function defocusing (QDA) and focusing (QFA) quadrupoles, and bending dipoles (D) are summarized in Table 2.

Table 2: MC SR magnet target parameters.

| Magnet | G (T/m) | B (T) | $D_x$ (cm) | $D_y$ (cm) |
| --- | --- | --- | --- | --- |
| Quadrupole QDA1/3 | -(31 - 35) | 9.0 | 2.8 | 0.3 - 0.5 |
| Quadrupole QFA2/4 | 85 | 8.0 | 3.6 - 5.6 | 0.2 |
| Bending dipole D | - | 10.4 | 3.7 - 4.8 | 0.2 - 0.4 |

The IR layout consistent with the machine parameters in Table 1 and vertical/horizontal beam size variations in the IR magnets are shown in Fig. 2. The final focus of muon beams in the IRs is provided by the quadrupole triplets formed by nine quadrupole magnets Q1-Q9. Six different apertures are considered to follow closely the beam sizes. The IR quadrupoles are divided into ~2-2.6 m long magnets to provide the necessary space for the tungsten masks in between. The bending dipole B1, placed directly after the final focusing (FF) triplet, generates a large dispersion function at the location of the sextupole nearest to the IP to compensate for the vertical chromaticity. It is composed of ~6 m long coils with tungsten masks in between. The IR magnet design parameters are summarized in Table 3. The aperture of the magnets is determined by the criterion $D_{x/y}=10\sigma_{max}+30$ mm.

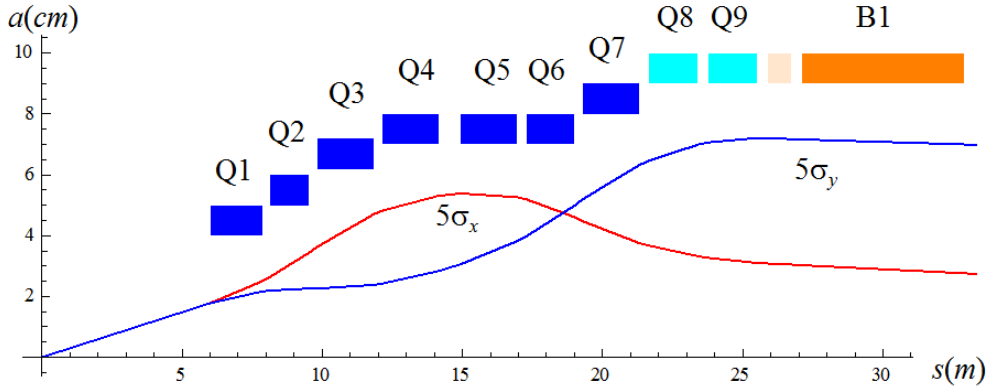

Figure 2: MC IR layout and beam size in magnets.

Table 3: IR Magnet Parameters.

| Parameter | Q1 | Q2 | Q3 | Q4-6 | Q7 | Q8-9 | B1 |
|---|---|---|---|---|---|---|---|
| Aperture (mm) | 80 | 100 | 124 | 140 | 160 | 180 | 180 |
| Operation gradient $G_{op}$ (T/m) | 250 | 200 | 161 | 144 | 125 | 90 | 0 |
| Operation field $B_{op}$ (T) | - | - | - | - | - | 2 | 8 |
| Magnet length (m) | 1.85 | 1.4 | 2.0 | 1.7 | 2.0 | 1.75 | 5.8 |

Previous analysis had shown that a horizontal shift of IR quadrupoles by ~1/10 of the aperture provides the dipole field component of ~2 T at the beam center and thus facilitates the chromaticity correction and dilutes background fluxes on the detector. The more round beams in the 3 TeV MC design do not allow for a significant shift of IR quadrupoles. Since the benefit of such a shift for detector backgrounds has not been demonstrated yet, it is not included in the present design. Another factor which is important for a TeV-scale MC is the radiation induced by neutrino fluxes [8]. In the IP nearest to the quadrupoles, the natural beam divergence is sufficient to spread it, but in more distant quadrupoles the additional bending field is necessary. Such bending field can be generated by special dipole coils.

**1.2 Towards 6 TeV MC**

The MC lattice design provides an average luminosity of up to $10^{35}$/cm$^2$/s at the top c.o.m. energy of 6 TeV [5]. The lattice allows a wide range of $\beta^*$ and muon energy variation necessary for the energy scan, provides an adequate protection of magnets and detectors from muon decay products and minimizes the severity of neutrino production in the arcs and in the IR and consequently in places where the neutrino approaches or exits the ground surface.

The protection of SC magnets and detector from radiation imposes strict limitations on the lattice choice, hence the design of the collider optics, magnets and MDI is closely intertwined. As a first approximation the IR design with $\beta^*$=3 mm was used, while for the arcs we will rescale the arc cell design of the 3 TeV collider.

The 1.5 TeV and 3 TeV muon collider designs assumed the presently available Nb$_3$Sn magnet technology. In the future the HTS technology can mature enough to become practical for large-scale applications, making it possible to consider 16 T pole tip fields in quadrupoles and 20 T fields in dipoles. Fig. 3 shows the 6 TeV MC interaction region design based on such assumption. The requirements on the IR quadrupole magnets are summarized in Table 4.

The dipole component in the IR quadrupoles may play a positive role by sweeping away from the detector the charged secondary particles created by decays in the incoming muon beam. To maximize this positive effect the strongest quadrupole in the FF multiplet, which is usually the second one from the Interaction Point (IP), should be defocusing.

Another important requirement for the FF multiplet is that the last quadrupole has to be also defocusing in order to reduce the "dispersion invariant" generated by the following strong dipole used in the chromaticity correction scheme discussed later. To satisfy both requirements simultaneously, the multiplet should be either a doublet or a quadruplet. The first option was used in the $E_{c.o.m.}$= 1.5 TeV design. For higher energies it is advantageous to use the second option which allows for much smaller $\beta^*$.

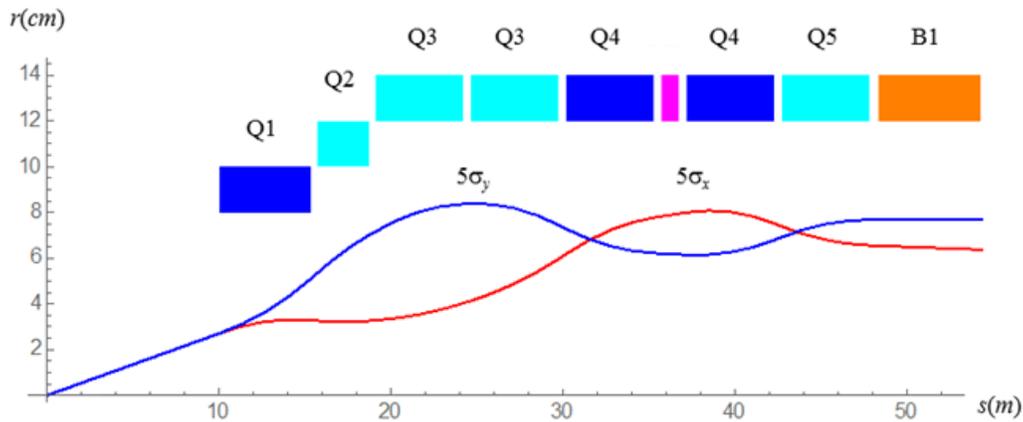

Figure 3: 6 TeV IR magnet apertures and $5\sigma$ beam envelopes for $\beta^* = 3$ mm and normalized emittance $\varepsilon\perp_N = 25\ (\pi)$ mm·mrad. Defocusing magnets are shown in cyan.

Table 4: 6 TeV IR Magnet Parameters

| Parameter | Q1 | Q2 | Q3 | Q4 | Q5 |
|---|---|---|---|---|---|
| Aperture (mm) | 160 | 200 | 240 | 240 | 240 |
| Operation gradient $G_{op}$ (T/m) | 200 | -125 | -100 | 103 | -78 |
| Operation dipole field $B_{op}$ (T) | - | 3.5 | 4.0 | 3.0 | 6.0 |
| Magnet length (m) | 5.3 | 3.0 | 5.1 | 5.1 | 5.1 |

## 2. Superconducting magnets

### 2.1 General considerations

The MC operating conditions present significant challenges to the SC magnets used in the Storage Ring and Interaction Regions. The IR magnets have to provide a high operating gradient and magnetic field in a large aperture to accommodate the large size of muon beams due to low

$β*$ as well as the cooling system to intercept the large heat deposition from the showers induced by decay electrons. The high level and distribution of heat deposition in the MC SR requires large aperture magnets to accommodate thick high-Z absorbers to protect the SC coils.

The high fields required for the MC SR and IR magnets call for advanced accelerator magnet technologies beyond traditional Nb-Ti magnets. The Nb-Ti superconductor used in all present accelerators has a critical temperature $T_{c0}$ of 9.8 K and an upper critical magnetic field $B_{c2}$ of 14.5 T, which limit the operation magnetic fields in accelerator magnets to 6 to 7 T at 4.5 K or 8 to 9 T at 1.9 K. A practical alternative to Nb-Ti is $Nb_3Sn$ superconductor with $T_{c0}$ of ~18 K and $B_{c2}$ of up to 30 T. Thanks to its superior properties, $Nb_3Sn$ enables magnetic fields in accelerator magnets up to 15-16 T at 4.5 K.

Progress in raising the performance parameters of commercial $Nb_3Sn$ superconducting composite wires in the late 1990s – early 2000s and impressive achievements of accelerator magnet technologies based on this superconductor in the past two decades make it possible to consider $Nb_3Sn$ accelerator magnets for the MC SR. Due to the significant challenges and uncertainties in operation conditions of magnets in the MC SR, it is reasonable to choose the nominal operation field on the level of 10-12 T, which provides a larger (up to 50%) operation margin.

The following sections present the conceptual designs and parameters of the bending arc magnets and IR magnets for a 3 TeV MC SR with an average luminosity of $4×10^{34}$ $cm^{-2}s^{-1}$ [11], [12]. The magnet coils are based on ~20 mm wide and ~2 mm thick multistrand Rutherford cables made of 1 mm diameter $Nb_3Sn$ strand. The larger Cable I is used for the SR dipoles and quadrupoles and the main IR quadrupole coils. The smaller Cable II is used for the auxiliary dipole coils in some quadrupoles.

**2.2 Storage Ring magnets**

The high level and distribution of heat deposition in MC SR require either large aperture magnets to accommodate thick high-Z absorbers to protect the SC coils, or an open mid-plane (OM) design to create a pass for the decay electrons to external absorbers. Despite the attractiveness of the OM approach, the analysis revealed serious issues for this magnet type. Besides the structural issues related to handling the large vertical forces in the coils with OM gaps, the dynamic heat load in the OM dipoles is still large even after implementation of appropriate protective measures. In the 1.5 TeV MC SR the dynamic heat load to cold mass in the OM dipoles is about 25 W/m. This is due to the decay electrons having a too large transverse momentum to pass through the open mid-plane with a strong vertical defocusing quadrupole field in the gap. Thus, a large aperture and an internal absorber would be needed even for the OM magnets. Furthermore, for muon beam energies of 1.5 TeV or higher, a dipole component is needed also in the quadrupoles to mitigate the neutrino radiation problem. This suggests using combined-function quadrupoles with additional dipole field components. Achieving the required level of both quadrupole and dipole components in OM combined-function magnets has also serious challenges.

Magnet apertures have to provide an adequate space for the internal absorber, vacuum insulation, magnet cold bore, and helium channel. The internal absorber geometry and size were estimated based on the MARS-calculated azimuthal distributions of heat deposition in the MC SR and a target attenuation factor of 100, to keep the average dynamic heat load in the SC magnets below 10 W/m. Taking into account that the quench limit of a $Nb_3Sn$ coil at 80% of its critical current is ~5 mW/g, and assuming a factor of 3 safety margin, the minimum absorber thickness to keep the heat deposition in the $Nb_3Sn$ coil below 1.5-1.7 mW/g is only ~2 cm. However, to keep the heat load in the MC SR magnets below 10 W/m the minimum absorber thickness increases to ~5 cm. This is a conservative assumption since the analysis was done without tungsten masks in between the magnets, which are very efficient in reducing the heat deposition in magnets.

The strong azimuthal dependence of the power density in the MC SR suggests using an absorber with a variable azimuthal thickness. In this study, an asymmetric elliptical absorber with 5 cm (inwards the ring) and 3 cm (outwards the ring) horizontal thicknesses was chosen. Assuming a round magnet aperture, its diameter is determined by the horizontal beam size of 56 mm and a total horizontal absorber thickness of 80 mm, 5 mm vacuum gaps, 2 mm thick Helium pipe, and 2 mm annular Helium channel on each side. The coil inner diameter (ID) in further analysis was rounded to 150 mm. The selected absorber size, geometry and magnet aperture have to be further optimized during the radiation heat deposition modeling. The beam center shift with respect to the coil center is ~10 mm. Thus, the radius of the good field quality region is 38 mm or a half of the coil inner radius. The free space between the elliptical absorber and the cold bore can be used for the absorber support and its cooling systems.

The magnets have 150 mm aperture and are based on 2-layer dipole and quadrupole coils, shown in Fig. 4. They have a cold iron yoke and operate at 4.5 K. The dipole (D) and quadrupole (Q) coils in the 4-layer combined-function magnets are powered independently. The D magnet provides a nominal operation field of 10.4 T. The combined Q/D magnet produces a dipole field of 8-9 T and a field gradient of up to 86 T/m. The arc magnets provide the operation fields and gradients with a large margin (within 65-85%) that can be used to optimize the space between dipoles for tungsten masks. All the low-order field harmonics in both magnet types, which include geometrical, coil magnetization and iron saturation components in the beam area, will be less than $10^{-4}$ of the main field component.

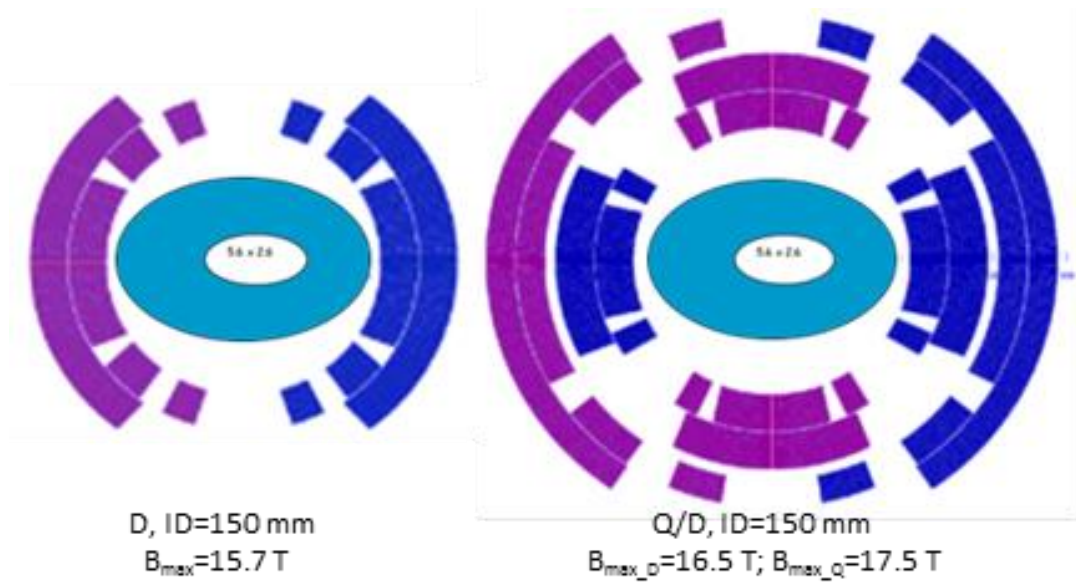

Figure 4: Bending dipole D (left) and combined function quadrupole-dipole Q/D (right) with the dipole coil outside of the quadrupole coil. The colors represent current direction in the coils.

## 2.3 Large-aperture IR magnets

The IR uses quadrupoles with 6 different apertures and nominal gradients, and a large-aperture IR dipole. Their cross-sections are shown in Fig. 5. The IR magnets are based on 2-layer shell-type coils and a cold iron yoke. The quadrupole apertures provide adequate space for the beam pipe, annular helium channel and the elliptical inner absorber (liner). Since the horizontal beam size is only ~60 mm, the dipole coil aperture is determined by the larger vertical beam size with an additional space for the cold beam pipe and helium channel. In a 3 TeV MC, it is of 180 mm, which allows fitting 4-5 mm thick slightly asymmetric beam absorbers inside the aperture, as it is done in MC arc magnets.

All the magnets operate at 4.5 K at 71-81% of the short sample limit (SSL) for Q1-Q9 quadrupoles and at 56% for the B1 dipole. As above, some fraction of the large operating margin in the B1 dipole can be used to increase the space between magnets for the tungsten masks. The 2 T dipole coils are nested in Q8-Q9 around the main quadrupole coils, and used to spread the neutrino flux. If necessary, they can be added to Q1-Q7 following a similar approach. The dipole coil cross-section in Q8-Q9 was optimized for operation with large margin (less than 75% of the SSL) to avoid coil quenching, simplify quench protection and provide a good dipole field quality. Total field harmonics including geometrical, coil magnetization and iron saturation in the IR magnets will be on the level of one unit ($10^{-4}$ of the main field) within 2/3 of the corresponding coil aperture (dark areas in Fig. 5).

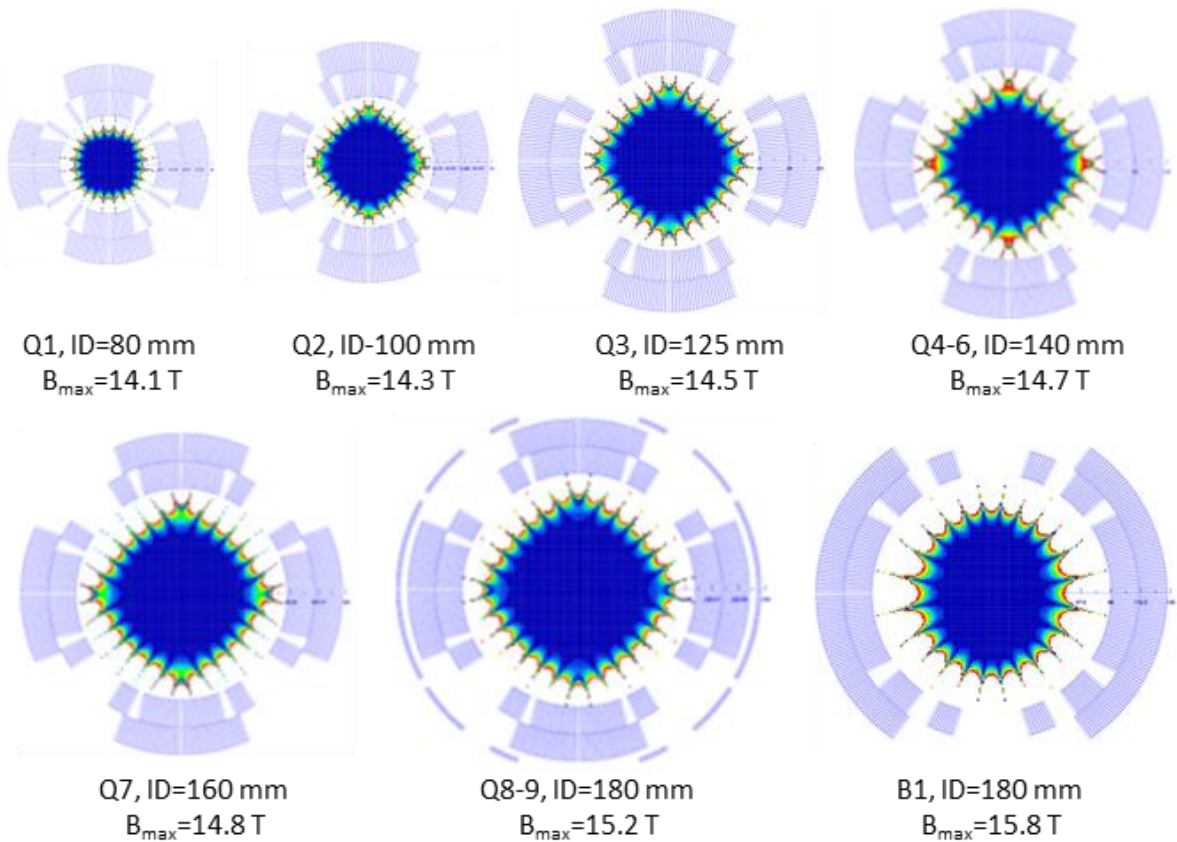

Figure 5: Coil cross-sections of IR quadrupoles Q1-Q9 and IR dipole B1.

## 2.4 Concluding remarks

Magnet coils, shown in Figs. 4 and 5, were designed to provide realistic field maps for the analysis and optimization of the arc lattice and IR design, as well as for studies of beam dynamics and protection of SC magnet coil against radiation. All these magnets are based on the traditional cos-theta coil geometry and $Nb_3Sn$ superconductor. The $Nb_3Sn$ magnet technology is new and still being developed for large accelerator facilities. However, the feasibility of Q1-Q7 quadrupoles, shown in Fig. 5, was demonstrated by testing series of short and up to 4-m-long $Nb_3Sn$ quadrupoles developed within the US-LARP program [15]. The cross-sections of these quadrupoles are shown in Fig. 6.

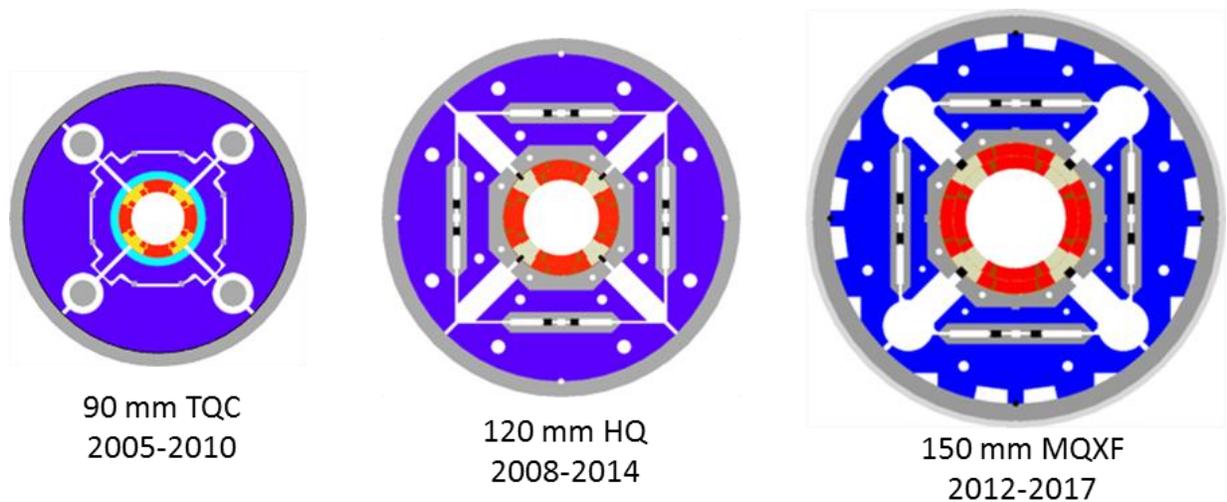

Figure 6: Large-aperture $Nb_3Sn$ quadrupoles developed and tested by US-LARP [16]-[18].

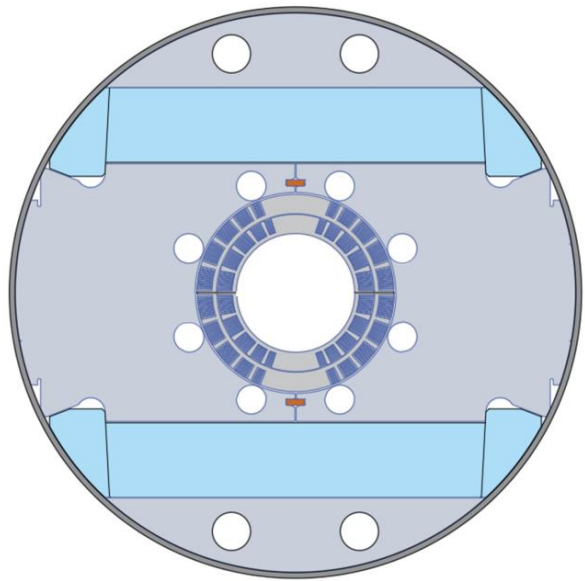

Figure 7: 120-mm aperture 11 T $Nb_3Sn$ dipole with stress management coil structure under development by US-MDP [20].

The high level of operating fields and the large apertures of the dipole and quadrupole magnets used in MC SR and IR lead to large Lorentz forces, which generate high mechanical stresses and strains in the superconducting coils. One of the major issues to be addressed in such magnets is stress management in the coil to avoid substantial degradation or even damage of the brittle $Nb_3Sn$ superconductor. Stress management concepts for brittle superconducting coils based on $Nb_3Sn$ as well as HTS superconductors, such as Bi2212 and REBCO, were proposed and are presently being developed by the US-MDP [19]. These concepts use a special metallic structure with radial shells and azimuthal bars to intercept and transfer the Lorentz forces inside the coil. Fig. 7 shows the 120-mm aperture 11 T $Nb_3Sn$ dipole with stress management cos-theta (SMCT) coil structure under development by the US-MDP [20]. The experimental studies and optimization of these concepts are still at a very early stage and need to be continued.

## 3. Radiation protection and Machine-Detector Interface

### 3.1 Earlier findings and new studies

Already in very early MC studies, it was found that:
- Photon, electron, positron and neutron fluxes and energy deposition in detector components are well beyond technological capabilities without reducing their values.
- Tungsten nozzles, starting at a few cm from the IP with a ±20 degree outer angle, are the most effective way (~1/500) of background suppression.
- The nozzles can also fully confine incoherent pairs if $B_{detector}$>3 T.
- High-field SC dipoles implemented in the final focus region, interlaced with quadrupoles and tungsten masks, provide further reduction of backgrounds.
- With such an IR design, the major source of backgrounds in a MC detector is muon decays in the region reduced to about ±25 m from the IP.
- Time gates promise substantial mitigation of the background problem in a MC detector.

These findings have recently been confirmed and the MC background problems further confronted in coherent studies by collider lattice and magnet designers, particle production and transport experts and detector groups [6], [7]. A consistent design now exists for a compact 1.5 TeV $\mu^+\mu^-$ collider ring, IR, MDI and chromatic correction section with large momentum acceptance and dynamic aperture, all based on high-field $Nb_3Sn$ SC magnets adequately protected against dynamic heat loads.

### 3.2 MARS modeling of backgrounds

Energy deposition and detector backgrounds are simulated with the MARS15 code [21]. All the related details of geometry, materials distributions and magnetic fields for lattice elements and tunnel in the ±200-m region from the IP, detector components, experimental hall and MDI are implemented in the model (Fig. 8). To protect the SC magnets and detector, 10 and 20 cm tungsten masks with $5\sigma_{x,y}$ elliptic openings in the IR magnet interconnect regions and sophisticated tungsten cones inside the detector were implemented (yellow in Fig. 8) into the model and carefully optimized. The 0.75 TeV muon beam is assumed to be aborted after 1000 turns. The minimal cutoff energies in this study range from 0.001 eV for neutrons, 0.1 MeV for electrons, positrons and photons, to 1 MeV for muons and charged hadrons. The cut-off energy in the tunnel concrete walls and soil outside is position-dependent and can be as high as a few GeV at 50-100 m from the IP compared to the minimal value in the vicinity of the detector.

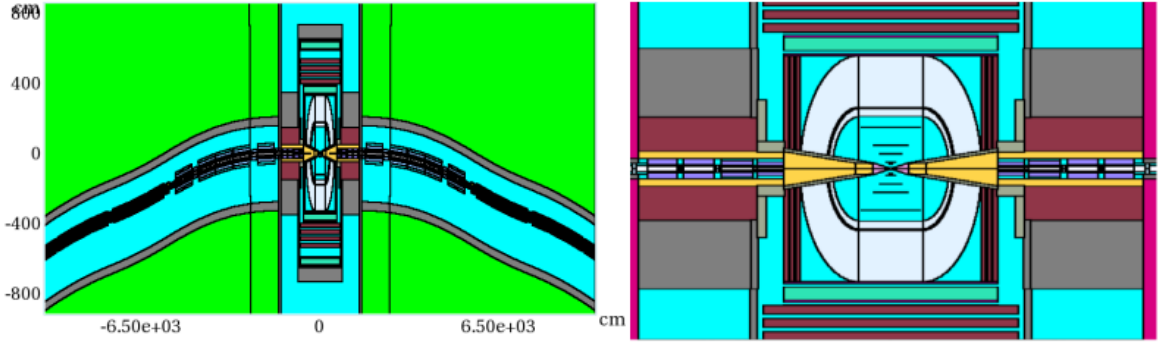

Figure 8: Plan views of MARS model for IR (left, axes in cm, |z|<10000 cm) and detector with MDI (right, |x|<500 cm, |z|<1500 cm).

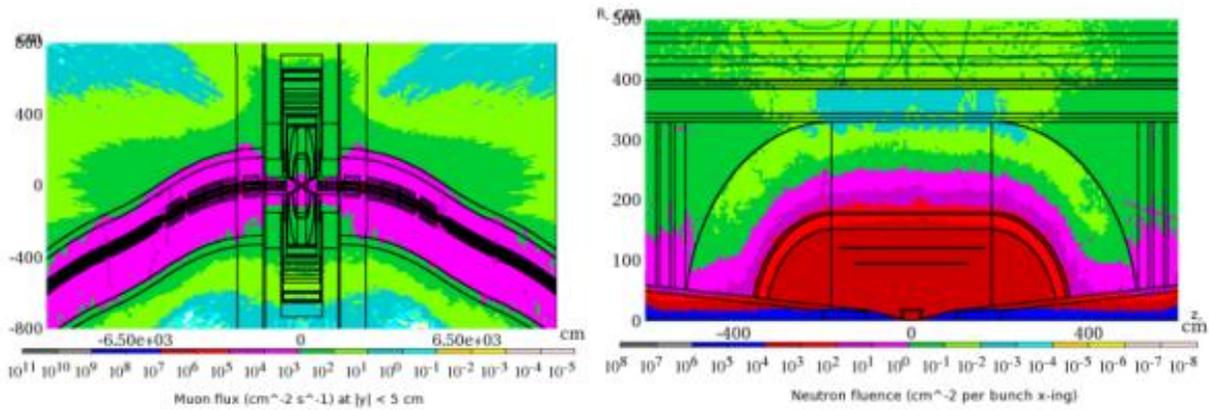

Figure 9: Muon isofluxes in IR (left) and neutron isofluences in the detector (right).

Fig. 9 (left) shows muon flux isocontours in the MC IR. These muons with energies of tens to hundreds of GeV illuminate the entire detector. They are produced in the Bethe-Heitler process by energetic photons from electromagnetic showers generated by decay electrons in the lattice components. The neutron isofluences inside the detector are shown in Fig. 9 (right). The maximum neutron fluence and absorbed doses in the innermost layer of the silicon tracker for a one-year operation are at a 10% level of that in the LHC CMS and ATLAS detectors at the nominal luminosity.

The dipoles close to the IP and tungsten masks in each interconnect region help substantially reducing background particle fluxes in the detector. The tungsten nozzles, assisted by the detector solenoid field, trap most of the decay electrons created close to the IP as well as most of incoherent $e+e-$ pairs from the IP. Their outer angle in the region closest to IP (6 to 100 cm) is the most critical parameter to optimize. The larger this angle the better background suppression, but the impact on the detector performance, especially in the forward region, becomes higher. The total numbers of photons and electrons entering the detector per bunch crossing are $1.5 \times 10^{11}$ and $1.4 \times 10^{9}$, respectively, for the minimal studied outer angle of the nozzle of 0.6 degrees, and reduced by three orders of magnitude for the MDI with an angle of 10 degrees (discussed further). Note that the inner opening shape and dimensions are carefully optimized in the nozzle region of 6<|z|<600 cm, and 5-cm thick borated polyethylene cladding is used at 100<z<600 cm where the outer angle is reduced to 5 degrees.

In the MARS15 runs, a source term for the detector simulations is calculated for all particles entering the detector through the MDI surface. This surface is defined around the IP (r=2.2 cm z=±13 cm), on the outer surface of the nozzle up to z=±600 cm, and on the inside of the detector at r=655 cm z=±750 cm. Corresponding high-statistics files have been generated for two 0.75-TeV muon beams, with minimal variation of particle weights, and with full information on the particle origin. The main characteristics of this source term are described in the next section. The source files are actively used by the European colleagues within the International Muon Collider Collaboration (IMCC)

### 3.3 Main characteristics of backgrounds

As found in earlier studies and confirmed with the current IR and MDI designs, the origin of all particles (except muons) entering the detector is the straight section of about ±25 m near the IP. The combined effect of angular divergence of secondary particles, strong magnetic field of the dipoles in the IR and tight tungsten masks in interconnect regions indicate that there is practically no contribution to non-muon detector backgrounds for z>±25 m. Excellent performance of the optimized nozzles and MDI shielding along with confinement of decay electrons in the aperture (forcing them to hit the nozzle on the opposite side of IP) result in the longitudinal distributions of particle origins with a broad maximum from 6 m to 17 m (IP side of the first dipole).

On the contrary, Bethe-Heitler muons hitting the detector are created in the lattice as far as 200 m from the IP, with 90% of them produced at ±100 m around the IP. As for all other particles, the fine structure of these distributions is related to the lattice details, with pronounced peaks always connected to the high-field SC dipole locations.

Most particles enter the detector through the nozzle outer surface as well as from backscattering of low-energy electrons, positrons and photons from the opposite-side nozzle tip. The maximum yield of photons and electrons is very close to the IP where the shielding is minimal. Neutron and charged hadron yields peak at z=±1 m. Muons enter the detector through the z=±750 cm plane (70%) and the nozzle (30%).

The fact that most particles (except muons) are produced close to the IP and are not affected by the strong dipole magnetic field results in the azimuthal symmetry of the source term with the corresponding distributions at MDI being flat. Angular distributions of these particles on the interface surface also have azimuthal symmetry. On the contrary, Bethe-Heitler muons are strongly affected by the magnetic fields of the dipoles on their long way to IP. As a result, azimuthal distributions of these muons have a strong asymmetry for the $\mu+$ beam. Secondary positive muons are deflected by the IR element magnetic fields to the same side as the positive muon bunch (negative horizontal direction). Secondary negative muons are deflected to the opposite side. Note that many muons hit the detector at large radii.

The total numbers of particles [7] entering the detector through the MDI surface along with particle mean momenta and energy flow show that the soft photons and neutrons are the major components. They are followed by electrons and positrons. Mean momenta of background particles are rather low, except for ~0.5 GeV/c charged hadrons and 22 GeV/c Bethe-Heitler muons. About 540 TeV of energy is brought to the detector by background particles per bunch crossing. Photons, neutrons and muons contribute about one third each to the energy flow, with two other components being small.

The momentum spread of particles entering the detector through the MDI surface is quite broad. With the kinetic cut-off energies indicated above used, photons and electrons with their ~MeV/c mean momenta, have always p<0.2 GeV/c. Hadron momentum reaches ~3 GeV/c, with very different mean values for neutrons and charged hadrons. Bethe-Heitler muons, illuminating the detector, do have the highest momenta of up to 200 GeV/c.

The time of flight (TOF) of background particles at the MDI surface has a significant spread with respect to the bunch crossing [7]. Two regions are clearly seen in the TOF distributions. The first one at TOF < 40 ns is related to the direct contributions from particles generated by muon beam decays in the ±17 m region not shielded by the strong magnetic field of the first dipole. The long tails for photons, electrons/positrons and neutrons are due to their bouncing and multiple interactions in the MDI components at low energies. The long tail for energetic Bethe-Heitler muons is associated with their production at large distances from the IP. These properties of the TOF distribution of the source term at MDI suggest that one can use timing in the detector to reduce the number of the readout background hits. The background neutron hit rate registered in the vertex and tracking silicon detectors can be reduced by a factor of several hundred by using the 7-ns time window.

### 3.4 Summary and extrapolation to higher energies

The recent developments in the design of Muon Collider SR and IR with the 1.5, 3 and 6 TeV c.o.m. energy based on $Nb_3Sn$ SC magnets along with substantial efforts in Monte-Carlo code developments and optimization studies of the machine-detector interface and appropriate detector technologies confirm that the severe background environment in such a challenging machine can be reduced to tolerable levels. The main characteristics of particle backgrounds entering the collider detector are studied in great details. A good understanding of background properties suggests the approaches to suppress the detector response to the non-IP related hits.

### 4. Conclusion

This White Paper brings together in a coherent form the results of the authors' previous studies on Muon Colliders and presents a design concept of the 6 TeV MC optics, the parameters of SC magnets, and a preliminary analysis of the protection system to substantially reduce radiation loads on the superconducting coils of the MC magnets as well as particle backgrounds in the collider detector. It also refines further the problem of neutrino fluxes generated by muon decays at a high-energy MC, corresponding deleterious effects induced by such neutrinos and their mitigation.

The SC magnets and detector protection considerations impose strict limitations on the lattice choice hence the design of the collider optics, magnets and MDI is closely intertwined. As the first approximation we use the Interaction Region (IR) design with $\beta^*=3$ mm, whereas for the arcs we rescale the arc cell design of the 3TeV collider.

The traditional cos-theta magnet coils were designed to provide realistic field maps for the analysis and optimization of the arc lattice and IR design, as well as for studies of beam dynamics and magnet protection against radiation. The high level of operating fields and large apertures of the SC magnets require using advanced superconductors. The major issue to address is the stress management in the coil to avoid substantial degradation or even damage of brittle SC coils. Stress management concepts for shell-type coils are being experimentally studied for high-field accelerator magnets based on LTS ($Nb_3Sn$) and HTS (Bi2212 and ReBCO) cables.

In the assumed IR design, the dipoles close to the IP and tungsten masks in each interconnect region (needed to protect magnets) help reducing background particle fluxes in the detector by a substantial factor. The tungsten nozzles in the 6 to 600 cm region from the IP (as proposed in the very early days of MC [2] and optimized later), assisted by the detector solenoid field, trap most of the decay electrons created close to the IP as well as most of the incoherent *e+e-* pairs generated in the IP. With sophisticated tungsten, iron, concrete and borated polyethylene shielding in the MDI region, a total reduction of background loads by about three orders of magnitude can be achieved.